# Thermal and Quantum Fluctuations Induced Additional Gap in Single-particle Spectrum of d-p Model

Partha Goswami

*Deshbandhu College, University of Delhi, Kalkaji, Delhi-110019, India.*

**Abstract.** The possibility of thermal and quantum fluctuations induced attractive interaction leading to a pairing gap $\Delta_{tq}$ in the single-particle spectrum of d-p model in the limit of a large N of fermion flavor is investigated analytically. This is an anomalous pairing gap in addition to the one with d-wave symmetry originating from partially screened, inter-site coulomb interaction. The motivation was to search for a hierarchy of multiple many-body interaction scales in high-$T_c$ superconductor as suggested by recent experimental findings. The pairing gap anisotropy stems from more than one sources, namely, nearest neighbour hoppings and the p-d hybridization but not the coupling of the effective interaction. The temperature at which $\Delta_{tq}$ vanishes may be driven to zero by using a tuning parameter to have access to quantum criticality only when $N \gg 1$. For the physical case $N = 2$, the usual coherent quasi-particle feature surfaces in the spectral weight everywhere in the momentum space below the pairing gap $\Delta_{tq}$. Thus it appears that the reduction in spin degeneracy has the effect of masking quantum criticality.

## 1. INTRODUCTION

There are a number of features of high temperature superconductors which do not completely fit within the single-band Hubbard model framework [1,2]. For example, the cuprate gap is set by the charge trans-

Corresponding author:Tel.91-1292439099;fax:91-1126449396; e-mail:physics_goswami @ yahoo.co.in

fer energy separating the copper and oxygen orbitals [3,4] as opposed to a Mott gap between copper d-states split by the on-site repulsion U. Furthermore, the metal-insulator transition (MIT) and superconductivity seems [5] to be difficult to explain in this framework at any finite U in two dimensions and higher. These were the reasons for considering a three-band Hubbard model [6,7](with spin degeneracy $N \geq 2$) to investigate MIT in an earlier work[8] ( hereinafter referred to as I ). A variant of the slave-boson (SB) mean field theoretic approach proposed by Kotliar and Ruchenstein (KR)[9] was applied in I. The advantages of the former over the latter were mentioned there. In the present work, which is a sequel to I, a novel d-fermion pairing mechanism leading to a gap $\Delta_{tq}$ in the single-particle excitation spectrum and the role played by the spin degeneracy vis-à-vis quantum criticality (QC) when this pairing instability sets in are reported. The motivation is to initiate the formulation of a multi-gap model for high-$T_c$ superconductor to explain the recent experimental observations [10-13] where (i) the pseudo-gap(PG) is reported to have two distinct energy scales(with a low-energy PG comparable to that of the superconductivity (SC) gap and a larger high-energy PG) and (ii) the possibility of multiple many-body interaction scales in high-$T_c$ superconductors is demonstrated. The p-and d-fermion nearest neighbor hoppings are introduced additionally keeping in view a previous report [14] in which it was shown that all the underlying d-p parameters have significant role. The specific aim here, of course, is to generate k-dependence of the onsite energies and to find their effect on nodal and anti-nodal gap contributions $\Delta_{tq}(N)$ and $\Delta_{tq}(AN)$. To capture the widely accepted d-wave gap anisotropy[2] in cuprates, the nearest-neighbor interaction counterpart in momentum space $U(\mathbf{k},\mathbf{q})$ is to be introduced [15] for transition from a momentum $\mathbf{q}$ to $\mathbf{k}$. This is taken to be separable in Ref.15 and is expanded in terms of basis functions corresponding to mixed symmetry states involving $d_{xy}$ and $d_{x^2-y^2}$. Once the singlet superconducting instability sets in, it will be shown that the effective gap involves the dominant one ($\Delta_{x^2-y^2}$) corresponding to pure $d_{x^2-y^2}$ wave and the sub-dominant $\Delta_{tq}$. The gaps ($\Delta_{xy}$, $\Delta_{tq}$) will be shown to be relevant for the normal phase.

As the first step to achieve the goals set above,the bose field fluctuations are integrated out completely taking partial trace in the grand partition function of the system for the metallic phase close to MIT to obtain an effective fermion Hamiltonian $H_{eff}$ in the momentum space. Due to weak attraction induced by thermal and $\lambda$-field fluctuations (the field $\lambda$ is responsible for enforcing the anti-commutativity of pseudo-fermion operators in the SB method in I and therefore the corresponding fluctuations are quantum-mechanical), it has been possible to show that the anomalous pairing of d-fermion field is nonzero. The corresponding gap ($\Delta_{tq}$) anisotropy in the single-particle excitation spectrum stems from more than one sources, namely, p- and d-fermion nearest neighbor hoppings ($t_{pp}$, $t_{dd}$) and the p-d hybridization but not the coupling of the effective interaction. Since the coupling involves thermal and quantum fluctuations both, the transition temperature may be driven to zero by tuning, say, the parameter 'u' in I to have access to quantum criticality (QC) only when the spin degeneracy $N \gg 1$ and the other gaps are zero. For $N = 2$, the access to QC is denied as then the attractive interaction would diverge. The usual coherent quasi-particle feature appearing in the spectral weight everywhere in the momentum space, for N=2, below the pairing gap is a confirmation of this fact. In the paired state the anti-nodal gap contribution $\Delta_{tq}(AN)$ monotonically increases with decreasing hole doping. The nodal gap contribution $\Delta_{tq}(N)$, expressible in the form $\Delta^2_{tq}(N) = \Delta^2_{tq}(AN) - 4(\sqrt{2}\, t_{dd} + t_{pp})^2$, also shows the same behavior. It is found that $T_{AN} > T_N$ where $T_{AN}$ and $T_N$, respectively, are the temperatures at which $\Delta_{tq}(AN)$ and $\Delta_{tq}(N)$ vanish. Furthermore, the characteristic energy of $\Delta_{tq}$ is found to be much greater than the superconducting (SC) gap energy.

The paper is organized as follows: In Sec.2 starting with the momentum space three-band Hubbard Hamiltonian (or d-p Hamiltonian) involving nearest neighbor (NN) hoppings and the bose field fluctuations, an effective fermionic Hamiltonian is obtained taking partial trace of the grand partition function. The Hamiltonian involves an attractive interaction which has its origin in the fluctuations of the field $\lambda$. In Sec.3 it is shown that this interaction leads to a novel pairing gap $\Delta_{tq}$ in the single-particle spectrum of the model under consideration in the limit of a large N of fermion flavor. The paper ends in Sec.4 with the reporting of a possible energy scale, corresponding to the gap $\Delta_{tq}$, about six times greater than the SC gap energy.

## 2. EFFECTIVE FERMION HAMILTONIAN

We start with the d-p model Hamiltonian in I involving boson fields ($\varphi, \psi$) and the field $\lambda$ enforcing the usual constraint of SB method. In momentum space the model Hamiltonian, involving bose mean-field values ($e_0, D_0$) (of $\varphi$ and $\psi$) and the corresponding fluctuations ($e_q, D_q$), can be written as (cf. Eqs.(6) and (7) in I)

$$H' = C_1 + H_{mean} + H_b^{(0)} + H_{b,f} \tag{1}$$

$$C_1 = N_s N \lambda (e_0^2 + D_0^2 - q_0) + N_s N U_d' D^2 \tag{2}$$

$$H_b^{(0)} = \lambda \sum_q e_q^\dagger e_q + (\lambda + U_d') \sum_q D_q^\dagger D_q \tag{3}$$

$$H_{mean} = \sum_{k\sigma} (\varepsilon_d^o + \lambda - \mu) d_{k\sigma}^\dagger d_{k\sigma} + \sum_{k\alpha\sigma} (\varepsilon_p^o - \mu) p_{k\alpha\sigma}^\dagger p_{k\alpha\sigma}$$

$$- \sum_{k\eta\sigma} (2i\, t\, \sin(k_\eta a/2)) \{ p^\dagger_{k\eta\sigma} (e_0 d_{k\sigma} + D_0 \mathrm{sgn}(\sigma) d^\dagger_{-k,-\sigma})$$

$$- (e_0 d^\dagger_{k\sigma} + D_0 \mathrm{sgn}(\sigma) d_{-k,-\sigma}) p_{k\eta\sigma} \} \tag{4}$$

$$H_{b,f} = \sum'_q \lambda_q \{ e_0 (e_{-q} + e_q^\dagger) + D_0 (D_{-q} + D_q^\dagger) \}$$

$$+ \sum'_q \lambda_{-q} \{ e_0 (e_q + e_{-q}^\dagger) + D_0 (D_q + D_{-q}^\dagger) \}$$

$$+ \sum'_{k\sigma q\eta} (1/\sqrt{N})(2it \sin(q_\eta a/2)) \{ (d^\dagger_{\pm q,\sigma} e_{\pm q-k} + \mathrm{sgn}(\sigma) D^\dagger_{\pm q+k} d_{\pm q,-\sigma}) p_{k\eta,\sigma}$$

$$- p^\dagger_{k\eta,\sigma} (e^\dagger_{\pm q-k} d_{\pm q,-\sigma} + \mathrm{sgn}(\sigma) d^\dagger_{\pm q,-\sigma} D^\dagger_{\pm q+k}) \} + \sum'_{q,k} (1/\sqrt{N}) \lambda_q (e^\dagger_{q+k} e_k +$$

$$D^\dagger_{q+k} D_k) + \sum'_{q,k} (1/\sqrt{N}) \lambda_q (d^\dagger_{k+q,\sigma} d_{k+q,\sigma}) \tag{5}$$

The $\varepsilon_d^o$ and $\varepsilon_p^o$, respectively, are the d- and p-fermion onsite energies plus the corresponding nearest neighbor hopping terms; $\mu$ is the chemical potential for fermion number. The hopping terms are given by $[-2t_{dd}(\cos k_x a + \cos k_y a)]$ and $[-4t_{pp} \sin(k_x a/2) \sin(k_y a/2)]$. The quantities ($e_0, D_0, \lambda, \mu$) can be determined from the mean field equations (Eqs.(8)-(11) in I). Here the primed summations correspond to $q \neq 0$. In what follows it will be shown that the last term in Eq.(5) yields the required attractive interaction. The Bose and Fermi parts are separated as far as possible in Eqs.(1)-(5). One may consider the grand partition function (GPF) $Z = \mathrm{Tr}\, \exp(-\beta H')$. In order to integrate out boson degrees of freedom the Hilbert space of the system may be factorized as $B \otimes F$ where B is the subspace on which boson operators act and F is the one on which fermion operators act. One can thus write $Z = Z_0 Z_f Z_{b-f}$ where $Z_0$ = Tr exp $(-\beta (C_1 + H_b^{(0)}))$, $Z_f$ = Tr exp $(-\beta H_{mean})$ and $Z_{b,f}$ = Tr exp $(-\beta H_{b,f})$. The assumptions of large spin degeneracy and proximity to MIT (where $e_0^2$ and $D_0^2 \to 0^+$) provides a $(1/N)$ expansion for GPF ($(1/N)$ is used as an expansion parameter to extrapolate to the physical case $N = 2$):

$$Z = Z_0 Z_f [ 1 + \sum (1/n!) \int d\tau_1 \int d\tau_2 ...... \int d\tau_n \langle T_t\{ H_{b,f}(\tau_1) H_{b,f}(\tau_2) ............... H_{b,f}(\tau_n) \} \rangle ] \qquad (6)$$

where all of the integrals extend over a finite domain $0 \leq \tau \leq \beta$, $1 \leq n \leq \infty$, and $\langle ............ \rangle$ refers to the thermal averages calculated with $H_b^{(0)}$. Though this equation is generally less useful than, say, an expression for thermodynamic potential [16] specially because of the difficulties associated with counting the disconnected graphs, it has been found quite adequate for arriving (the details will be reported elsewhere) at an expression for effective fermion Hamiltonian as the approximate exponentiation of connected graph contributions is possible by those of disconnected graphs; one has to assume that bose momenta should not be larger than Fermi momenta for this purpose. As bose fields merely correspond to book-keeping devices to keep track of the immediate environment surrounding fermions in a slave-boson formalism, the assumption is not in conflict with the methodology adopted.

The first order contribution to GPF is zero which is quite clear from the structure of terms in $H_{b,f}$. The relevant second-order contributions ($\tilde{A}, \check{A}$) are noted below:

$$\tilde{A} \approx (\beta^2/2N) \sum'_{qk\sigma} \lambda^2_q \{ \langle d^\dagger_{k+q,\sigma} d^\dagger_{-(k+q),-\sigma} \rangle d_{-k,-\sigma} d_{k\sigma} + \langle d_{-k,-\sigma} d_{k\sigma} \rangle d^\dagger_{k+q,\sigma} d^\dagger_{-(k+q),-\sigma} + d^\dagger_{k+q,\sigma} d_{k+q,\sigma} \} \quad (7)$$

$$\check{A} \approx -(\beta^2/2N) \sum'_{k\sigma q\eta} (4t'^2 \sin^2(q_\eta a/2)) \{ \langle d^\dagger_{\pm q,\sigma} d_{\pm q,\sigma} \rangle p^\dagger_{k\eta\sigma} p_{k\eta\sigma} + d^\dagger_{\pm q,\sigma} d_{\pm q,\sigma} \langle p^\dagger_{k\eta\sigma} p_{k\eta\sigma} \rangle \}$$

$$- (\beta^2/2N) \sum'_{k\sigma q\eta} (4t''^2 \sin^2(q_\eta a/2)) \{ d^\dagger_{\pm q,\sigma} d_{\pm q,\sigma} + p^\dagger_{k\eta\sigma} p_{k\eta\sigma} \} . \qquad (8)$$

Here ($t'^2$, $t''^2$) are given by the expressions

$$t'^2 = t^2 \{ \coth \beta \lambda / 2 - \coth \beta (\lambda + U_d')/2 \} \qquad (9)$$

$$t''^2 = t^2 [ \exp(\beta(\lambda + U_d'))/ (\exp(\beta(\lambda + U_d')) - 1) - \exp(\beta \lambda)/(\exp(\beta \lambda) - 1) ]. \qquad (10)$$

As will be seen below, the thermal averages in Eqs.(7) and (8) will be determined in a self-consistent manner. The hyperbolic and exponential functions in (9) and (10) correspond to the normal boson propagators $\langle e^\dagger_q e_q \rangle$ and $\langle D^\dagger_q D_q \rangle$. The anomalous boson propagators have not been considered above as these are zero to the leading order in (1/N). The exponentiation of the contributions in (7) and (8) by the third and higher order terms in (6) leads to the effective fermion Hamiltonian $H_{eff}$. One obtains

$$H_{eff} = C + H_{mean} + H_f \qquad (11)$$

where $C = C_1 + O(1/N)$ and $H_f = -(\tilde{A} + \breve{A})/\beta$. The interaction corresponding to the term $(-\tilde{A}/\beta)$ is attractive and has its origin in the fluctuations of the field $\lambda$ enforcing the anticommutativity of pseudo-fermion operators introduced in the present slave-boson technique to account for the effect of the p-d hybridization. This interaction term, where the quantum fluctuations play an important role together with the thermal energy $k_B T$, is responsible for the singlet pairing instability (SPI) to be discussed shortly. In the zero temperature limit, the interaction is expected to be finite only when $N \gg 1$. The other term $(-\breve{A}/\beta)$, which also arises from the p-d hybridization and the bose field fluctuations, corresponds to repulsive one and therefore does not favor this instability. In fact, this repulsion supports a collective mode. The corresponding natural frequencies are given by the poles of the density response function of the system. The behavior of the collective mode near the SPI is currently under investigation.

## 3. ANOMALOUS PAIRING

In this section it will be shown how the attractive interaction obtained above leads to a gap in the single-particle spectrum. To this end. the thermal averages in $H_{eff}$ (Eqs.(7) and (8)) may be evaluated in a self-consistent manner introducing the Green's functions

$$G^{(a)}_{dd}(k,\sigma,\tau;-k,-\sigma,0) = -\langle T\{d_{k,\sigma}(\tau) d_{-k,-\sigma}(0)\}\rangle$$

$$G_{pd\eta}(-k,-\sigma,\tau;-k,-\sigma,0) = -\langle T\{p^\dagger_{-k\eta,-\sigma}(\tau) d_{-k,-\sigma}(0)\}\rangle \qquad (12)$$

etc. For the Fourier transform of the anomalous function $G^{(a)}_{dd}(k,\sigma,\tau;-k,-\sigma,0)$, in particular, one obtains

$$G^{(a)}_{dd}(\acute{K},z) \approx \Delta(\acute{K})/\Gamma(\acute{K},z)$$

$$\Gamma(\acute{K},z) = [(z-(\varepsilon^o_d+\lambda-\mu))(z-(\varepsilon^o_p-\mu)) - \sum_\eta 4t^2(e_0^2+D_0^2)\sin^2(k_\eta a/2) - \Delta^2(\acute{K})], \qquad (13)$$

where $z = i(2n+1)\pi/\beta$ and $\acute{K} = (k,\sigma)$. The gap $\Delta(\acute{K})$ is given by

$$\Delta(\acute{K}) = - \sum{'}_q (\beta/2N) \lambda^2_q \{ \langle d_{k+q,\sigma} d_{-(k+q),-\sigma} \rangle + \langle d_{k-q,\sigma} d_{-(k-q),-\sigma} \rangle \} . \qquad (14)$$

The pairing gap is anisotropic in momentum space and odd function of $\acute{K}$. Usually the anisotropy stems from the fact that coupling of the effective interaction being in a specific angular momentum state other than the s-wave one. However, the anisotropy here has its origin in the nearest neighbor hoppings ($t_{pp}, t_{dd}$) and p-d hybridization. Furthermore, one finds that the insulator-to-metal transition discussed in I(signaled by the mean field values of the bose fields $\varphi$ and $\psi$, respectively, corresponding to empty and doubly occupied sites acquiring non-zero values) is followed by the onset of singlet pairing instability (SPI) as the temperature is lowered in the near zero doping limit. The transition temperature can be driven to zero, tuning the parameter 'u' in I, to have access to quantum criticality provided the spin degeneracy N is strictly much greater than one. The quantum fluctuations then get the full opportunity to dominate over thermal fluctuations.

The situation when the spin degeneracy is reduced to N=2 is worth investigating for obvious reason. It may be seen easily that in this case the transition temperature could not be driven down to zero as then the attractive coupling approaches infinity leading to a total catastrophe. Thus the reduction in spin degeneracy has the effect of masking quantum criticality. One obtains unconventional gap equation for N=2. The departure from the conventional BCS result prima facie lies in the facts that (i) the poles of the temperature function Fourier transform $G^{(a)}_{dd}(\acute{K}, z)$ are given by

$$\varepsilon_j = (1/2)(\varepsilon^o_d + \lambda + \varepsilon^o_p \pm E_k), j = (1,2) \qquad (15)$$

$$E_k^2 = [(\varepsilon^o_p - (\varepsilon^o_d + \lambda))^2 + 16t^2(e_0^2 + D_0^2) \sum{'}_{k\eta} \sin^2(k_\eta a/2) + \Delta^2(k)] \qquad (16)$$

$$\Delta(k) \approx - \sum{'}_q (\beta/2N) \lambda^2_q \{ \langle d_{k+q\uparrow} d_{-k-q\downarrow} \rangle + \langle d_{k+q\downarrow} d_{-k-q\uparrow} \rangle \} \qquad (17)$$

which is form-wise much more complex from the corresponding result $\pm\sqrt{(\varepsilon_k^2 + \Delta_{sc}^2)}$ in BCS theory (where $\Delta_{sc}$ is k-independent) and (ii)the k-dependence of the gap, which gets determined here by solving the gap equation and the mean field equations of the unknowns ($e_0, D_0, \lambda, \mu$) simultaneously, stems from more than one sources, namely, p- and d-fermion hoppings [$-2t_{dd}(\cos k_x a + \cos k_y a)$] and $[-4t_{pp} \sin(k_x a/2) \sin(k_y a/2)]$ and the p-d hybridization $[t^2 (e_0^2 + D_0^2) \sum'_{k\eta} \sin^2(k_\eta a/2)]$ but not the coupling of the effective interaction. As mentioned above, the possibility of mixed symmetry states involving $d_{xy}$ and $d_{x^2-y^2}$ and the corresponding gap anisotropy exist for N=2 provided unscreened, inter-site interaction term for d-fermions is also included in the model. It is being hoped that the bulk of the cuprate physics would then get captured successfully. The mean field equations are given by (cf. Eqs. (8)-(11) in I with $q_0 = 1/2$)

$$(e_0^2 + D_0^2) = (1/2) - (1/N_s) \sum_k (u_k^2 f(\varepsilon_1) + v_k^2 f(\varepsilon_2)) + F$$

$$F = -(1/(N N_s)) \sum_q \langle e^\dagger_q e_q + D^\dagger_q D_q \rangle$$

$$\lambda = (1/N_s) \sum_k \{4t^2 F_k^2 / E_k\} (f(\varepsilon_1) - f(\varepsilon_2))$$

$$\lambda + U_d' = (1/N_s) \sum_k \{4t^2 F_k^2 / E_k\} (f(\varepsilon_2) - f(\varepsilon_1))$$

$$(1 + \delta)/4 = (1/(2N_s)) \sum_k (f(\varepsilon_1) + f(\varepsilon_2) + f(\varepsilon_p^0)) + D^2$$

$$F_k^2 = 2(\sin^2(k_x a/2) + \sin^2(k_y a/2))$$

$$\Delta(k) \approx \sum'_q (\beta/N) \lambda^2_q (\Delta(k)/E_k)(f(\varepsilon_1) - f(\varepsilon_2)). \tag{18}$$

Here $N_s$ is the number of unit cells. The Bogoliubov coherence factors $(u_k^2, v_k^2)$ are given by

$$u_k^2 = (1/2) \{1 + ((\varepsilon^o_p - \varepsilon^o_d - \lambda)/E_k)\},$$

$$v_k^2 = (1/2) \{1 - ((\varepsilon^o_p - \varepsilon^o_d - \lambda)/E_k)\}. \tag{19}$$

In equations above, $f(\varepsilon_1), f(\varepsilon_2)$ and $f(\varepsilon_p^0)$ are Fermi functions. One adds a hole on the oxygen site for hole doping ($\delta > 0$) and therefore $f(\varepsilon_p^0) \neq 0$. For electron doping ($\delta < 0$), since one removes a hole from the Cu-site, $f(\varepsilon_p^0) = 0$. As in I, the momentum integrals involved in the mean field equations above have been evaluated with constant density of states. It may be noted that the fluctuations driven pairing mechanism reported here does not require phonon/exciton mediation.

## 4. DISCUSSION AND CONCLUSION

One of the central issues in a theory of strongly correlated systems is the existence or not of well-defined quasi-particles. This question is best addressed by studying the spectral weight (SW) $A(k, \omega)$. In a Fermi liquid (FL), the SW is dominated by quasi-particle peaks. One finds from (15) that, for N=2, the peaks are at $\omega = (\varepsilon_1, \varepsilon_2)$. One also finds that the insulator-to-metal transition investigated in I is followed by the onset of singlet pairing instability (SPI) as the temperature is lowered in the zero doping limit. Both the metallic and the ordered phases are characterized by the formation of the quasi-particle peaks in $A(k,\omega)$. However, since quantum fluctuations lead to SPI, one may expect the instability to correspond to a quantum critical point (QCP) transition and a possible FL picture breakdown in this case. On the contrary, we find that the ordered phase has coherent quasi-particle scenario (i.e. perfect FL picture) everywhere in the momentum space below the pairing gap. This is not surprising, since for N=2 quantum fluctuations never get the full opportunity to dominate over thermal fluctuations and the character of the transition remains normal.

The analysis of photoemission experiments [12,17] in cuprates has generated considerable debate [18, 19] over the concept of energy gaps with distinct doping dependence. For the gap $\Delta_{tq}$ under consideration, the anti-nodal gap contribution $\Delta_{tq}(AN)$ monotonically increases with decreasing hole doping. The nodal gap contribution, expressible in the form

$$\Delta^2_{tq}(N) = \Delta^2_{tq}(AN) - 4(\sqrt{2}\, t_{dd} + t_{pp})^2, \qquad (20)$$

also shows the same behaviour. Equation (13) in I is used to arrive at this conclusion. It is also found that $T_{AN} > T_N$, where $T_{AN}$ and $T_N$ are the temperatures at which $\Delta_{tq}(AN)$ and $\Delta_{tq}(N)$ vanish. The dominant hopping term $t_{pp}$ in (20) is 0.6 eV. One finds that the characteristic energy of $\Delta_{tq}$ is also of the same order which is larger than the SC gap energy of 0.1eV.

We have demonstrated above a novel mechanism of anomalous pairing leading to a high-energy gap in the single-particle spectrum of the d-p model in addition to the well-known gap with d-wave symmetry. It must, however, be emphasized that our approach is rigorous only in the limit of a large N of fermion flavor.